\documentclass[pdflatex,sn-vancouver-num]{sn-jnl}

\usepackage{graphicx}%
\usepackage{multirow}%
\usepackage{amsmath,amssymb,amsfonts}%
\usepackage{amsthm}%
\usepackage{mathrsfs}%
\usepackage[title]{appendix}%
\usepackage{xcolor}%
\usepackage{textcomp}%
\usepackage{manyfoot}%
\usepackage{booktabs}%
\usepackage{listings}%

\usepackage[ruled]{algorithm2e}

\theoremstyle{thmstyleone}%
\theoremstyle{thmstyletwo}%

\theoremstyle{thmstylethree}%

\begin{document}

\title[Article Title]{Quest for a clinically relevant medical image segmentation metric: the definition and implementation of Medical Similarity Index.}


\author*[1]{\fnm{Szuzina} \sur{Fazekas}}\email{fazekas.szuzina@semmelweis.hu}

\author[1]{\fnm{Bettina Katalin} \sur{Budai}}
\author[1]{\fnm{Viktor} \sur{Berczi}}
\author[1]{\fnm{Pal} \sur{Maurovich-Horvat}}

\author[1,2]{\fnm{Zsolt} \sur{Vizi}}\email{zsvizi@math.u-szeged.hu}

\affil*[1]{\orgdiv{Medical Imaging Centre}, \orgname{Semmelweis University}, \orgaddress{\street{Üll?i str. 78a}, \city{Budapest}, \postcode{1085}, \state{Pest}, \country{Hungary}}}

\affil[2]{\orgdiv{Bolyai Institute}, \orgname{University of Szeged}, \orgaddress{\street{Aradi vértanúk tere 1.}, \city{Szeged}, \postcode{6720}, \state{Csongrád-Csanád}, \country{Hungary}}}


\abstract{\textbf{Background:} In the field of radiology and radiotherapy, accurate delineation of tissues and organs plays a crucial role in both diagnostics and therapeutics. While the gold standard remains expert-driven manual segmentation, many automatic segmentation methods are emerging. The evaluation of these methods primarily relies on traditional metrics that only incorporate geometrical properties and fail to adapt to various applications. This study aims to develop and implement a clinically relevant segmentation metric that can be adapted for use in various medical imaging applications.

\textbf{Methods:} Bidirectional local distance was defined, and the points of the test contour were paired with points of the reference contour. After correcting for the distance between the test and reference center of mass, Euclidean distance was calculated between the paired points, and a score was given to each test point. The overall medical similarity index was calculated as the average score across all the test points. For demonstration, we used myoma and prostate datasets; nnUNet neural networks were trained for segmentation.

\textbf{Results:} An easy-to-use, sustainable image processing pipeline was created using Python. The code is available in a public GitHub repository along with Google Colaboratory notebooks. The algorithm can handle multislice images with multiple masks per slice. Mask splitting algorithm is also provided that can separate the concave masks. We demonstrate the adaptability with prostate segmentation evaluation.

\textbf{Conclusions:} A novel segmentation evaluation metric was implemented, and an open-access image processing pipeline was also provided, which can be easily used for automatic measurement of clinical relevance of medical image segmentation.}


\maketitle

\section{Background}

Medical image segmentation plays a crucial role in numerous clinical applications, particularly in radiation treatment planning. Modern radiotherapy enables the precise delivery of high radiation doses for the target volume while minimizing the exposure of the surrounding healty tissues (organs-at-risk). Accurate delineation of both the tumor and the surrounding organs-at-risk is essential, however this remains a very time consuming and also subjective task \citep{sherer2021}, leading to significant variability among clinicians. Moreover, uncertainity and inconsistency in target volume delineation represent the the largest error throughout the radiotherapy workflow - from treatment planning to the delivery process \citep{segedin2016}. The quality of radiation protocol has been shown to heavily influence patient outcome. Deficiencies in the treatment plan are reported to result in significantly reduced survival, with head and neck cancer patients experiencing two-year decrease in overall survival \citep{peters2010}. Significant intra- and interobserver variability in the segmentation of target volumes in breast cancer has been associated with negative impact in treatment outcome \citep{buelens2022}. Accurate brain tumor segmentations are also vital as these have direct impact on surgical planning \citep{warfield2004}. 

In response to these challenges, there are an emerging number of automated segmentation methods have been developed. In the field of radiology, nowadays more and more artificial intellgience-based solutions are in clnical practice \citep{fazekas2022}. However, the evaluation of such algorithms remains inconsistent due to the lack of standardized assessment protocols. Widely used metrics - such as traditional area-based and distance-based metrics - only incorporate geometrical properties and often fail to adapt to specific clinical context. 

Low correlation between segmentation metrics and dosimetric changes for OARs was shown in brain tumor patients \citep{poel2021}. Poel et al. investigated overall 23 different metrics, including similarity measures (Dice, Jaccard, AUC, etc.), distance measures (Hausdorff, AHD, etc.) and classical measures (sensitivity, specificity, etc.). They found all the metrics have limited predictive value for treatment quality and consequently suggested revision towards clinically oriented metrics.

\section{Materials and Methods}

For reproducibility, detailed description of the used methods are provided, along with the available GitHub repository \citep{szuzina_bld} containing the codes. The work was carried out in three main stages: data collection, coding and segmentation evaluation. 

The data collection consisted of two datasets: our Institution's fibroid segmentation dataset and a freely available prostate anatomic segmentation dataset. The former one was used for demonstrating the pipeline, while the purpose of the latter one was to show the adaptability of the workflow. We chose six-six patients for the two datasets: two easy, two moderate and two difficult cases, and we used these images for testing the nnUNet neural network, in that way test masks for six-six patients' MRI images were generated.

We programmed the calculation of Medical Similarity Index in Python programming language. The code is freely available in a GitHub repository and an executable Google Colaboratory notebook is provided, which contains all the necessary code for the evaluation. We provide simple solutions for contour pairing and mask splitting, along with other calculation steps, but all these ideas can be modified according to the users' needs or ideas.

The segmentation evaluation was conducted on the six-six selected patients' generated test masks of the two datasets. The traditional metrics were calculated as well as MSI with different settings (\texttt{il} and \texttt{ol} parameters).

\subsection{Datasets}
The image processing pipeline was developed and fine-tuned using pelvic MRI image segmentations of uterine fibroids. This segmentation dataset includes various mask shapes and scenarios, as uterine fibroids can be very diverse in number and appearance. As we used the results of not fine-tuned neural networks, we could test how the pipeline performs on challenging use-cases.

Patients who underwent uterus artery embolization in the Semmelweis University Medical Imaging Center between May 2016 and September 2020 were selected (overall 161 patients), and the pre-treatment baseline MR images were chosen. Overall, 31 patients were excluded, from which 10 patients' DICOM images were damaged, 16 patients had non-contourable fibroids (due to an extreme number of fibroids or non-identifiable fibroid boundary), and five patients were excluded because of the presence of adenomyosis.

All the MRI examinations were conducted at the Semmelweis University Medical Imaging Centre on 1.5 T equipment (Philips Ingenia 1.5T, Philips Healthcare, Best, Netherlands). A routine contrast-enhanced pelvic MRI protocol was applied, which included T1W, T2W, T2W-SPAIR, and contrast-enhanced T1W-SPIR sequences. The imaging was executed with a 90° flip angle, 80 s echo time, 0.70 mm voxel spacing, and 400×400 reconstruction matrix.

The MRI images were exported from the institutional PACS (Picture Archiving and Communication System) in DICOM format and converted to NIfTI format. A radiology resident manually segmented the T2W sequence using 3DSlicer software (slicer.org), which were then validated by two expert radiologists with more than 10 years of experience in pelvic MRI imaging.

Out of the 130 patients, 124 were used for training and 6 were selected for testing: 2 easy, 2 moderate and 2 difficult cases. The test segmentations were generated using the DKFZ \texttt{nnUNet} framework \citep{isensee2021}. We created a \texttt{nnUNet} usage tutorial in the Google Colaboratory notebook, which handles all the necessary steps before the preprocessing and training of the \texttt{nnUNet} framework. Only the Google Drive links of the training and testing zip files need to be provided. The \texttt{nnUNet} is widely used in medical imaging as it is an automated framework so it can configurate its hyperparameters based on the dataset fingerprint. The executable Google Colaboratory notebook for the neural network learning is also provided for reproducibility. Specifically, we used 2D U-Net with the default planner for 5 folds, 100 epochs per fold, the initial learning rate was 0.01. The training was done on an NVidia GeForce RTX 3060 12GB Dual V2 OC video card, one epoch took about 100 seconds.

An open-access multi-site dataset for prostate MRI segmentation was used for demonstration purposes with corresponding anatomical prostate segmentations \citep{liu2021}. Prostate segmentation on MRI images is a challenging task due to the heterogeneity of prostate structure. However, in the case of radiotherapy, precise prostate anatomic segmentation is crucial due to the proximity of the urinary bladder. The segmentation of the prostate is important not only for radiotherapy, but enables to follow up the volume of the prostate in the disease progression, helps multimodal image registration, designate the region of interest for computer aided diagnosis (CAD) and contribute to the staging of prostate cancer (using PI-RADS) \citep{nai2020}. In this application area, important organs are very close to the segmentation area, which makes the evaluation very crucial. When radiation therapy is applied, any outer deviation of the segmentation mask can cause radiation damage to the urinary bladder. This damage will have a huge impact on living conditions such as incontinence.
\label{prostate} 

The dataset consists of a total of 115 prostate T2W MRI images and corresponding segmentation masks. The images are from six different data sources out of three public datasets (NCI-ISBI 2013 \citep{bloch2015}, I2CVB \citep{lemaitre2015}, PROMISE12 \citep{litjens2014}). The preprocessing included conversion to NIfTI format, centering the prostate, and resizing to a size of 384×384 in the axial plane.

From the 115 MRI scans, 109 were used for training and 6 were selected for testing: 2 easy, 2 moderate and 2 difficult cases. The training was done with nnUNet using the same Google Colaboratory notebook as in case of the fibroid dataset.

\subsection{Definition of Medical Similarity Index}

The Medical Similarity Index (MSI) addresses the problems arising with the traditional segmentation evaluation metrics. This metric is based on a bidirectional local distance to evaluate the similarity of two contours, based on the work of \cite{kim2015}. In the case of image segmentation, we can assume that the point set is discrete. The reference contour is considered the gold standard, and we measure the similarity of the test contour to this reference. The reference and test contours are paired based on \textit{bidirectional local distance} (denoted by $\mathrm{BLD}$), giving a score to each test point. The final $\mathrm{MSI}$ is calculated by the average of these scores along all test points.

Firstly, we define the minimal distance between a point $\boldsymbol{p}$ and a contour $C$ (discrete point set) as
\begin{equation}
	\mathrm{d_{min}}(\boldsymbol{p}, C) = \min_{\boldsymbol{q} \in C} \| \boldsymbol{p} - \boldsymbol{q} \| _2 .
\end{equation}
For calculating the \textit{forward minimum distance} ($\mathrm{FMinD}$), the closest point of the reference contour from a given test point must be found, i.e.
\begin{equation}
	\mathrm{FMinD}(\mathrm{\boldsymbol{p}_{test}}, R) =  \mathrm{d_{min}}(\mathrm{\boldsymbol{p}_{test}}, R).
\end{equation} \label{fmind}

For computing the \textit{backward maximum distance} ($\mathrm{BMaxD}$), we iterate through all the reference points and find the closest test points for each. If there exists a reference point for which the endpoint of this distance is the previously selected test point, then the maximum of these distances will be chosen as $\mathrm{BMaxD}$, i.e. for the reference contour $R$ and a test point $\mathrm{\boldsymbol{p}_{test}}$, we have:
\begin{equation}
	\mathrm{BMaxD}(R, \mathrm{\boldsymbol{p}_{test}}) =  \max_{ \mathrm{\boldsymbol{p}_{r}} \in R} \Big\{ \mathrm{d_{min}}(\mathrm{\boldsymbol{p}_{r}}, T): \| \mathrm{\boldsymbol{p}_{r}} - \mathrm{\boldsymbol{p}_{test}} \| _2 = \mathrm{d_{min}}(\mathrm{\boldsymbol{p}_{r}}, T) \Big\}.
\end{equation} \label{bmaxd}

The $\mathrm{BLD}$ (corresponding to one test point) is the maximum of $\mathrm{FMinD}$ and $\mathrm{BMaxD}$, for a point $\mathrm{\boldsymbol{p}}$ and a contour $R$:
\begin{equation}
	\mathrm{BLD}(\mathrm{\boldsymbol{p}_{test}}, R) = \max\Big\{\mathrm{FMinD}(\mathrm{\boldsymbol{p}_{test}}, R), \mathrm{BMaxD}(R, \mathrm{\boldsymbol{p}_{test}})\Big\}.
\end{equation} \label{bld}

The \textit{signed} $\mathrm{BLD}$ is negative if the test point is inside the reference contour and positive if the point is outside, indicated as $\mathrm{BLD}^{\pm}$.

\begin{figure}
	\centering
	\includegraphics[width=13 cm]
	{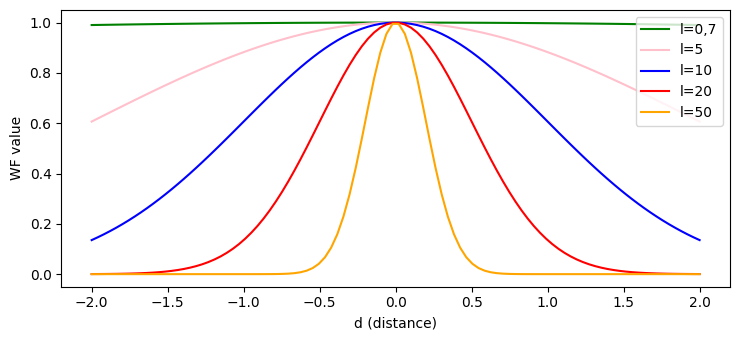}
	\caption{Weight Function. The $\mathrm{WF}(d, \texttt{l})$ weight function with different fixed $\texttt{l}=0.7, 5, 10, 20, 50$ level values.}
	\label{fig:wf}
\end{figure}

The \textit{Medical Similarity Index} (shortly $\mathrm{MSI}$) is calculated based on a modified Gaussian curve, we call it the \textit{Weight Function}, denoted by $\mathrm{WF}(d, \texttt{l})$. Different \textit{Weight Function} curves are demonstrated in Fig. \ref{fig:wf}, using different \texttt{l} values.

\begin{equation}
	\mathrm{WF}(d,\texttt{l}) = exp \left( -\frac{d^2}{2 \cdot (10/\texttt{l})^2} \right).
\end{equation} \label{wf}

The value $\texttt{l}$ is a user-defined constant, which can also have different values with respect to the test point is either inside or outside the reference contour. This penalty level can differentiate between the inside and outside test contour deviation, so the test contour's inside or outside alteration can be scored differently. 

The score is calculated with the value $\texttt{il}$ if the test point is inside the reference contour: 
\begin{equation}
	\mathrm{MCF_i}(\boldsymbol{p}, R) = \mathrm{WF}(\mathrm{BLD}^{\pm}(\boldsymbol{p}, R), \texttt{il}),
\end{equation}
\(\text{where } \texttt{il} \text{ is pre-defined constant.}\)
The score is calculated with the value $\texttt{ol}$ if the test point is outside the reference contour:
\begin{equation}
	\mathrm{MCF_o}(\boldsymbol{p}, R) =  \mathrm{WF}(\mathrm{BLD}^{\pm}(\boldsymbol{p}, R), \texttt{ol}),
\end{equation}
\(\text{where } \texttt{ol} \text{ is pre-defined constant.}\)
The definition of the inside and outside penalty levels can reflect the particular needs of the clinical application. If the outside deviation has serious consequences in the current medical circumstances, i.e. a vital organ is close to the segmented lesion, the outside penalty can be higher. If the inside deviation is unacceptable for the current medical use, i.e. calculating the tumor volume for radiation therapy, the inside level can be set to a greater value. Using these user-defined constants in the $\mathrm{MCF}$, a score is assigned to each test point. The final $\mathrm{MSI}$ value is the average of all scores along all test points. If the test point $\boldsymbol{p}$ is inside the reference contour, the $\mathrm{MCF_i}$ formula is used, if $\boldsymbol{p}$ is outside of the reference contour, $\mathrm{MCF_o}$ is used, with the following notations:

\begin{equation}
	\mathrm{I}(R) = \{ \boldsymbol{x}: \boldsymbol{x} \text{ is inside the reference contour}\}
\end{equation}
and
\begin{equation}
	\mathrm{O}(R) = {\overline{\mathrm{I}(R)} \setminus R},
\end{equation}
where $\overline{\mathrm{I}(R)}$ denotes the complement of $\mathrm{I}(R)$

The final $\mathrm{MSI}$ is defined by the following formula for a test contour $T$ and a reference contour $R$:
\begin{equation}
	\mathrm{MSI}(T, R) = \frac{1}{n} \left( \sum_{\substack{\boldsymbol{p} \in T, \\ \boldsymbol{p} \in \mathrm{I}(R)}} \mathrm{MCF_i}(\boldsymbol{p}, R) + \sum_{\substack{\boldsymbol{p} \in T, \\ \boldsymbol{p} \in \mathrm{O}(R)}} \mathrm{MCF_o}(\boldsymbol{p}, R) \right),
\end{equation} \label{msi}

where $n = |T|$. Clearly, if the reference and test contours intersect, the $\mathrm{MCF}$ score of the intersection points is zero.

\subsection{Traditional metrics for evaluating segmentation}

The demonstrated pipeline also includes calculating the most commonly used traditional image segmentation metrics: Dice score, Jaccard score, and average Hausdorff distance. We implemented each metric with a deterministic approach, since the number of points in our calculations does not suggest the use of randomized implementations (as in widely used implementations such as \texttt{SciPy} \citep{taha2015}).

\begin{figure}
	\centering
	\includegraphics[width=13 cm]
	{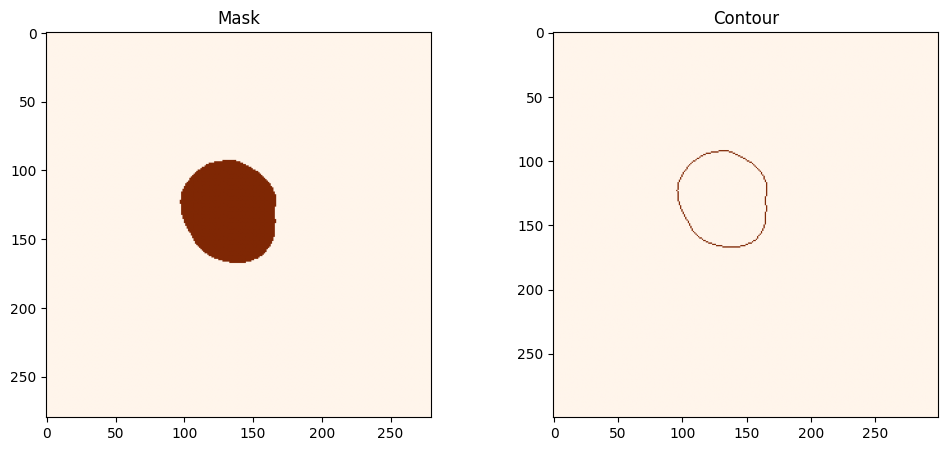}
	\caption{Definition of mask and contour. We define a mask as all the pixels corresponding to the segmented area, while we mean the boundary of the mask by the contour.}
	\label{fig:mask_contour}
\end{figure}

An image can be defined as an $m \times n$ matrix. We define a mask as a group of the pixels (i.e. matrix elements), which indicates the region of interest. We interpret a contour as the boundary of a mask, as clearly shown in Fig. \ref{fig:mask_contour}. 

One of the most commonly used image segmentation metrics is the \textit{Dice index}. 

\begin{equation}
	D(M_R, M_T) = \frac{2\cdot|M_R \cap M_T |}{|M_R| + |M_T|},
\end{equation}
\(\text{where } M_R \text{ is the reference mask, } M_T \text{ is the test mask}, \\ |M_R| \text{ is the (set) cardinality of the mask.} \)
\vspace{0,3 cm}

The Sorensen-Dice coefficient originates from statistics, where the similarity of two sets was assessed. The Dice score is also called \textit{F1 score} and can be calculated as the harmonic mean of precision and recall.

\begin{equation}
	\mathrm{F_1}(M_R, M_T) = \frac{2 \cdot \mathrm{TP}(M_R, M_T)}{2 \cdot \mathrm{TP}(M_R, M_T) + \mathrm{FP}(M_R, M_T) + \mathrm{FN}(M_R, M_T)}    
\end{equation}

where $\mathrm{TP}$ denotes the number of true positive, $\mathrm{FP}$ the false positive, $\mathrm{FN}$ false negative elements (or pixels) \citep{muller2022}, where
\begin{itemize}
	\item the true positive pixels mean the marked pixels which are correctly marked;
	\item the true negative pixels are the non-marked pixels in the ground truth segmentation which are not marked by the proposed segmentation;
	\item the false positive pixels are signed by the test segmentation, but correspond to the background on the reference segmentation.
\end{itemize}

The \textit{Jaccard score} is also an area-based metric, which can be defined as the ratio of the number of elements of the intersection of two sets divided by the number of elements of the union of the two sets.
\begin{equation}
	J(M_R, M_T) = \frac{|M_R \cap M_T|}{|M_R \cup M_T|}
\end{equation}
In image processing Jaccard score is known as IoU (Intersection over Union).

The \textit{Hausdorff distance} is a widely used metric in medical image analysis, which measures the largest segmentation error. 
\begin{equation}
	d_H(M_R, M_T) = \max \Big\{\sup_{\boldsymbol{x}\in M_R} d(\boldsymbol{x}, M_T), \sup_{\boldsymbol{y}\in M_T} d(M_R,\boldsymbol{y}) \Big\},
\end{equation}
\( \text{where } d(\boldsymbol{x}, M_T) \text{ is defined as } d(\boldsymbol{x}, M_T) = \inf_{\boldsymbol{y} \in M_T} d(\boldsymbol{x}, \boldsymbol{y})\)

The directed average Hausdorff distance from point set $X$ to point set $Y$ can be calculated as the sum of all minimum distances from all points in $X$ to $Y$ divided by the number of points in $X$. The average Hausdorff distance is given by the average of the directed average Hausdorff distance from point set $X$ to point set $Y$ and from $Y$ to $X$ \citep{aydin2021}. 
In our implementation, we paired the reference and test contours before the calculation of average Hausdorff distance. In some special cases it may differ from the usual implementation, but it is a more intuitive and logical approach. For each reference point the closest test point is selected, so if a test point of another contour is closer to the current reference point than the closest test point of its test contour pair, our implementation will use the latter one, while the usual implementations use the former one. 

\subsection{Practical issues with contours: contour pairing, mask splitting} \label{split}

\begin{figure}
	\centering
	\includegraphics[width=13 cm]
	{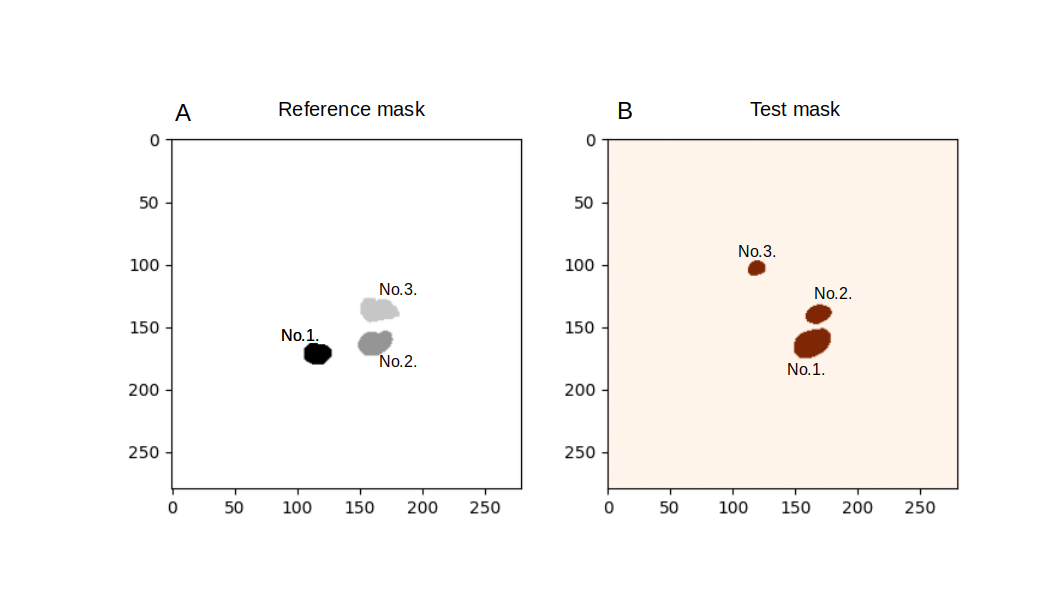}
	\caption{Representative slice for contour pairing. In this representative slice, the contour pairing with the closest center of mass method can not be done. For reference contour no. 1, the closest test COM is test contour no. 1. For reference contour no. 2, the closest test COM is also test contour no. 1. For this reason the algorithm can not handle this slice.}
	\label{fig:pairing_exc}
\end{figure}

In some medical applications, a segmentation mask consists of only one segment (e.g., prostate anatomic segmentation). In other cases, such as the fibroids, there can be more than one contour on one image slice. In such cases, the reference and test contours must be paired for reasonable metric calculations — not only for MSI but also for average Hausdorff distance. In our implementation, we calculated the center of mass (COM) for each reference and test contours, and assign the closest test COM for each reference COM. If there are more than one test COM assigned for a reference COM (as in Fig. \ref{fig:pairing_exc}), we consider the slice as a special case, which needs manual intervention.

In image recognition, object detection is a well known discipline with variety of publications and algorithms \citep{amit2021}. In this field, a commonly observed challenging task is to handle overlapping objects. In medical imaging, two lesions or organs can touch each other without overlap - there are no pixels which correspond to both objects. For this reason the general mask splitting algorithms can not be used in our pipeline. In our segmentation masks, if two contours are touching, it means that some pixels of one contour is adjacent to some pixels of the other contour. In that case, the two contours are considered as one mask (consequently one contour), but the further calculations would need them as separate contours.

For selecting the masks for splitting, it is possible to set a minimum mask area (\texttt{MIN\_AREA}), only the masks above this threshold will be candidates for splitting. The used algorithm (see Algorithm \ref{fig:algo}.) separates the concave masks based on the ratio of the mask area and the convex hull area.  A user defined threshold ratio is set (the default value is 1.2), only the masks above this ratio will be separated. The algorithm finds the convexity defects of the concave masks, which can be defined as the maximal distance of the masks from the corresponding side of the convex hull. Based on further parameters, the cut will be applied along the maximal convexity defect points.

\begin{algorithm}
	Calculate the area of the mask\;
	\If{area of the mask $> \texttt{MIN\_AREA}$}{
		Find the contour of the mask\;
		Find the convex hull\;
		\If{there are more than 2 convex hull points and $\texttt{area\_ratio} > 1.2$}{
			Find the convexity defects\;
			\If{length of convexity defect $< 0.5 \cdot \texttt{max\_length}$}{
				Throw out
			}
			\If{only one convexity defect point is left}{
				we do not discuss: 'horseshoe' shaped mask
			}
			\If{there are more convexity defect points left}{
				the splitting is between the two closest ones
			}
		}
	}
	\caption{The steps of the mask splitting algorithm.}
	\label{fig:algo}
\end{algorithm}

The $\mathrm{OpenCV}$ \citep{OpenCV} package was used for the mask splitting algorithm, by name the \texttt{findContours}, \texttt{convexHull} and \texttt{convexityDefects} functions.

\section{Results}

\subsection{Demonstration of the image processing pipeline}

\begin{figure}
	\centering
	\includegraphics[width=13 cm]
	{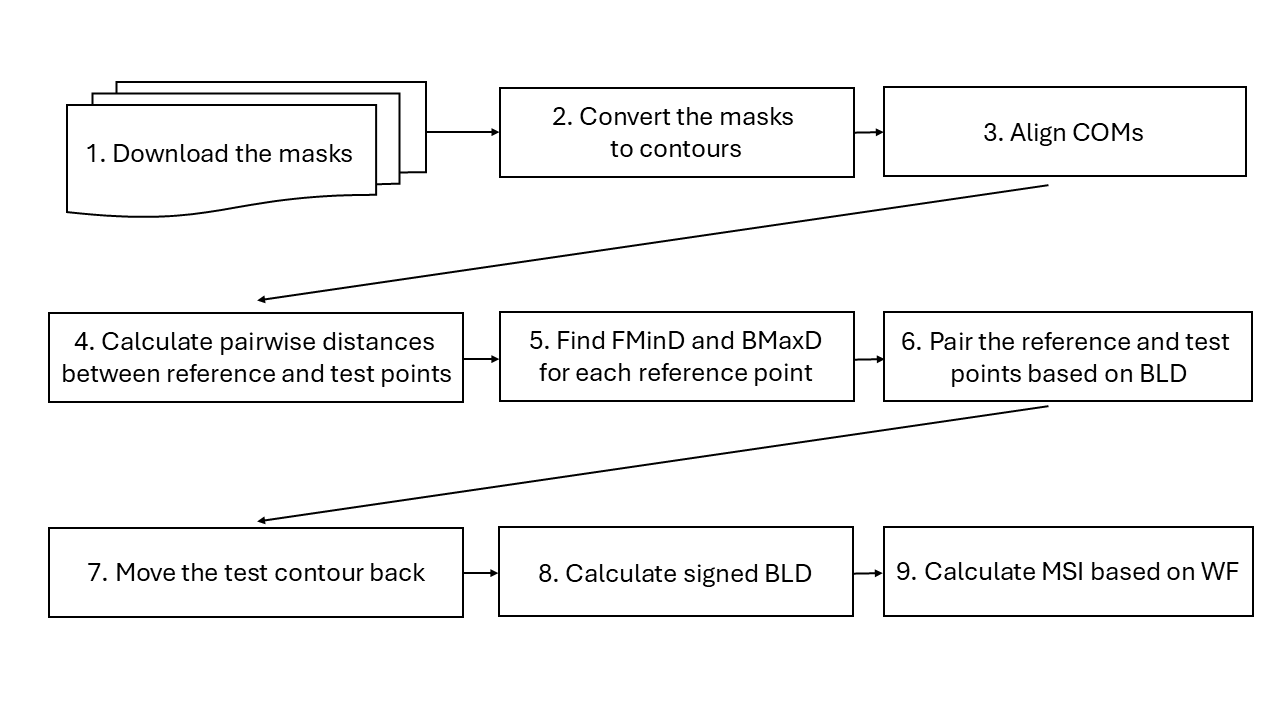}
	\caption{Flowchart of our pipeline. The calculation steps consist of (1) downloading the image masks, (2) converting the masks to contours for further analysis, (3) align the center of masses of the reference and test masks, (4) calculate the pairwise distances between all the reference and test points, (5) find the $\mathrm{FMinD}$ and $\mathrm{BMaxD}$ for each of the reference points, (6) pair the reference and test points based on $\mathrm{BLD}$, (7) move the test contour back to the original location, (8) calculate the distances between the paired points in the original location and finally (9) calculate the final $\mathrm{MSI}$ value based on the $\mathrm{WF}$.}
	\label{fig:flow1}
\end{figure}

The created Google Colaboratory notebook provides a user-friendly tutorial: it guides the user through all the necessary steps to calculate, visualize, and experiment with the MSI. See details in the Appendix. We demonstrate the created pipeline with pelvic MRI fibroid segmentation masks, since these masks provide a variety of shapes and mask arrangements with many different mask layouts and serious mask deviations. Our purpose was to present the adaptability of the pipeline, that is the reason why we used a not fine-tuned neural network for segmentation, to produce as many segmentation errors and deviations as possible. The calculation code executes the steps summarized in Fig. \ref{fig:flow1}.

\begin{figure}
	\centering
	\includegraphics[width=13 cm]
	{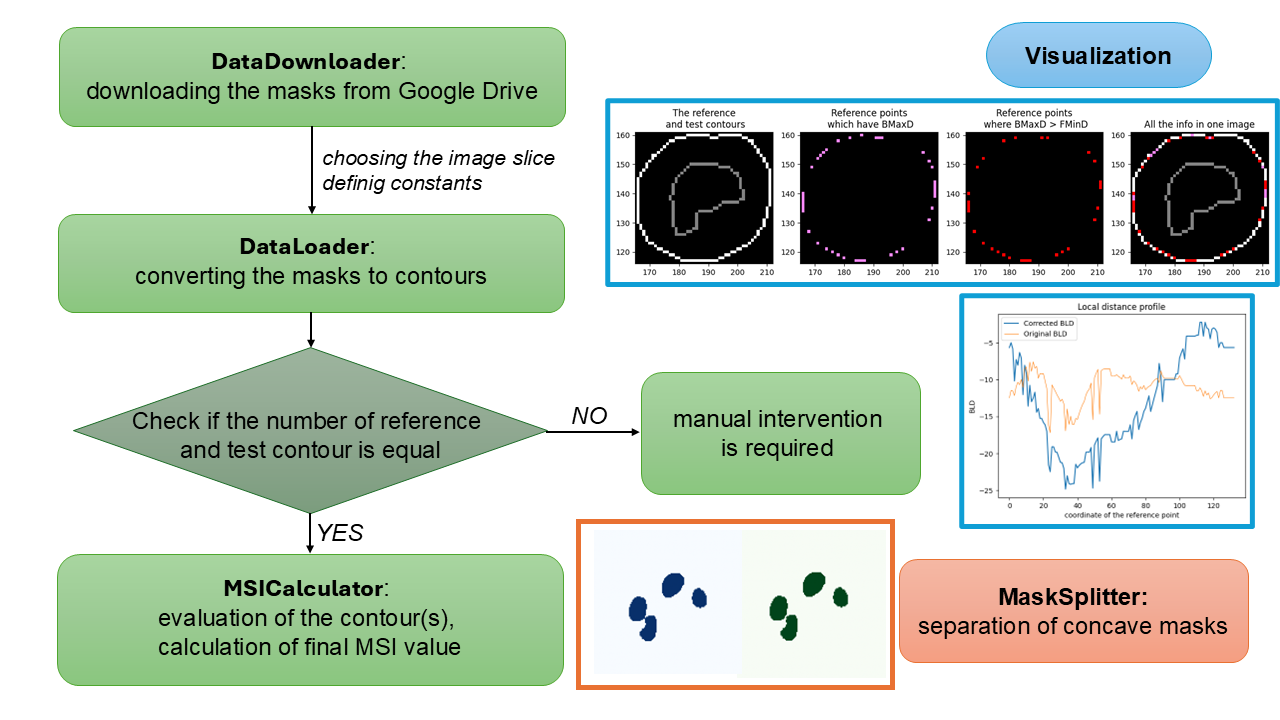}
	\caption{Flowchart of the runnable notebook. The notebook downloads the segmentation masks from the provided Google Drive link. After the image slice needs to be chosen for further analysis, and the user defined parameters can be modified. The masks are converted into contours, and the number of reference and test masks is compared. If there are the same number of reference and test masks, the automatic evaluation can be continued. If not, mask splitting or manual intervention is needed. The automatic evaluation will result in the final MSI value, as well as other figures for visualization.}
	\label{fig:flow2}
\end{figure}

An attached Jupyter notebook gives the opportunity for the user to evaluate image masks. The outline of the evaluation steps in the notebook is given in Fig. \ref{fig:flow2}.

\subsection{The pipeline separates the problematic slices}

The mask splitting algorithm can separate the problematic slices, as it works only with the filtered masks. The filtering is based on mask size and the ratio of mask area and convex hull area (see section \ref{split}). The splitting of these masks is enabled for the user, or any other manual intervention can be done with this set of filtered masks.

\begin{figure}
	\centering
	\includegraphics[width=13 cm]
	{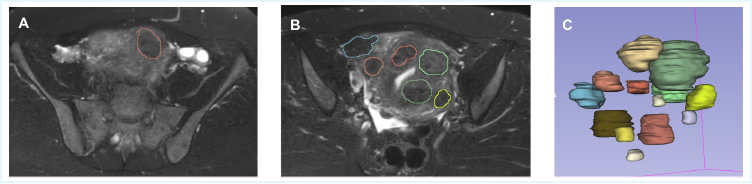}
	\caption{Representative slices of a patient with 14 fibroids. In our dataset, the number of fibroids in one patient can vary from 1 to 14. In case of the patient with 14 fibroids, the number of masks in one slice ranges from 1 to 6. In some slices (see Panel A.), there is only one mask, but other slices may interfere with more fibroids, thus there are slices with six masks as well (see Panel B.). The T2W SPAIR axial MRI images with the segmentation masks, as well as the 3D reconstruction of the masks are shown in the images.}
	\label{fig:numbers_on_slice}
\end{figure}

In the demonstration concerning fibroids, there are many of these problematic slices as the number of fibroids in one patient has a wide range (from 1 to 14 in our dataset), see an example on Fig. \ref{fig:numbers_on_slice}.

\begin{figure}
	\centering
	\includegraphics[width=13 cm]
	{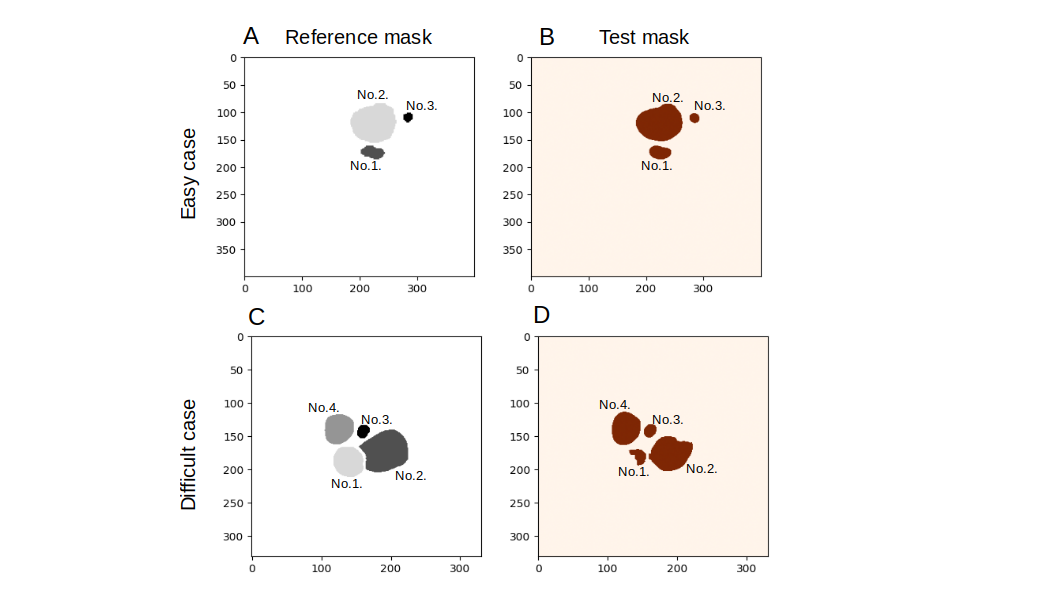}
	\caption{Examples for contour pairing algorithm. The contour pairing algorithm can pair easy and difficult cases as well. The reference masks (Panel A. and Panel C.) are colored by gray, the test masks (Panel B. and Panel D.) are colored by red.}
	\label{fig:pairing}
\end{figure}

The contour pairing is done via searching for the closest center of mass method. The algorithm works properly on easy and difficult slices as well (see examples in Fig. \ref{fig:pairing}). However, there are some special configurations, when the closest COM method does not work.

\begin{figure}
	\centering
	\includegraphics[width=13 cm]
	{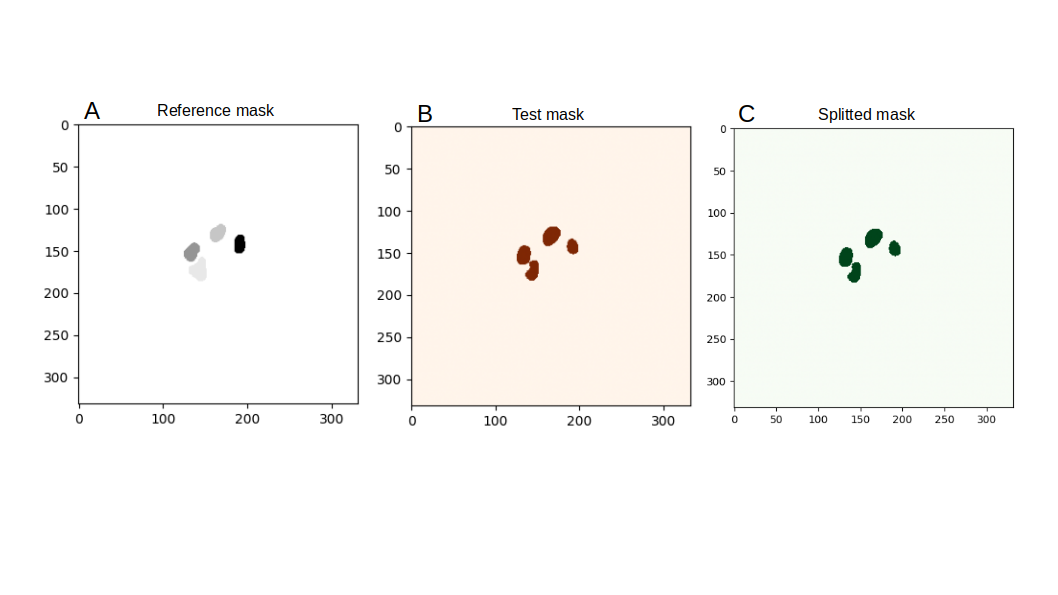}
	\caption{Representative slice for mask splitting. In this representative slice, there are four reference contours (Panel A.), while there is only three separate test contours (Panel B.), as two masks are touching. With the mask splitting algorithm, it is possible the separate the touching masks (Panel C.) and the automatic evaluation can be performed.}
	\label{fig:split_figure}
\end{figure}

The mask splitting algorithm was prepared to solve the challenging cases where two masks are touching (see Fig. \ref{fig:split_figure}). If the contour pairing is not possible due to the different number of reference and test masks, mask splitting might be able to solve this problem. If there is one or more contours, which has common pixels (so they are touching) and they fulfill the criteria for the splitting algorithm, the separation of the previously mentioned contours enables the algorithm to pair the contours and the user can continue the evaluation with the pipeline.

\subsection{The hyperparameters of MSI can be modified according to the user's needs}

\begin{figure}
	\centering
	\includegraphics[width=13 cm]
	{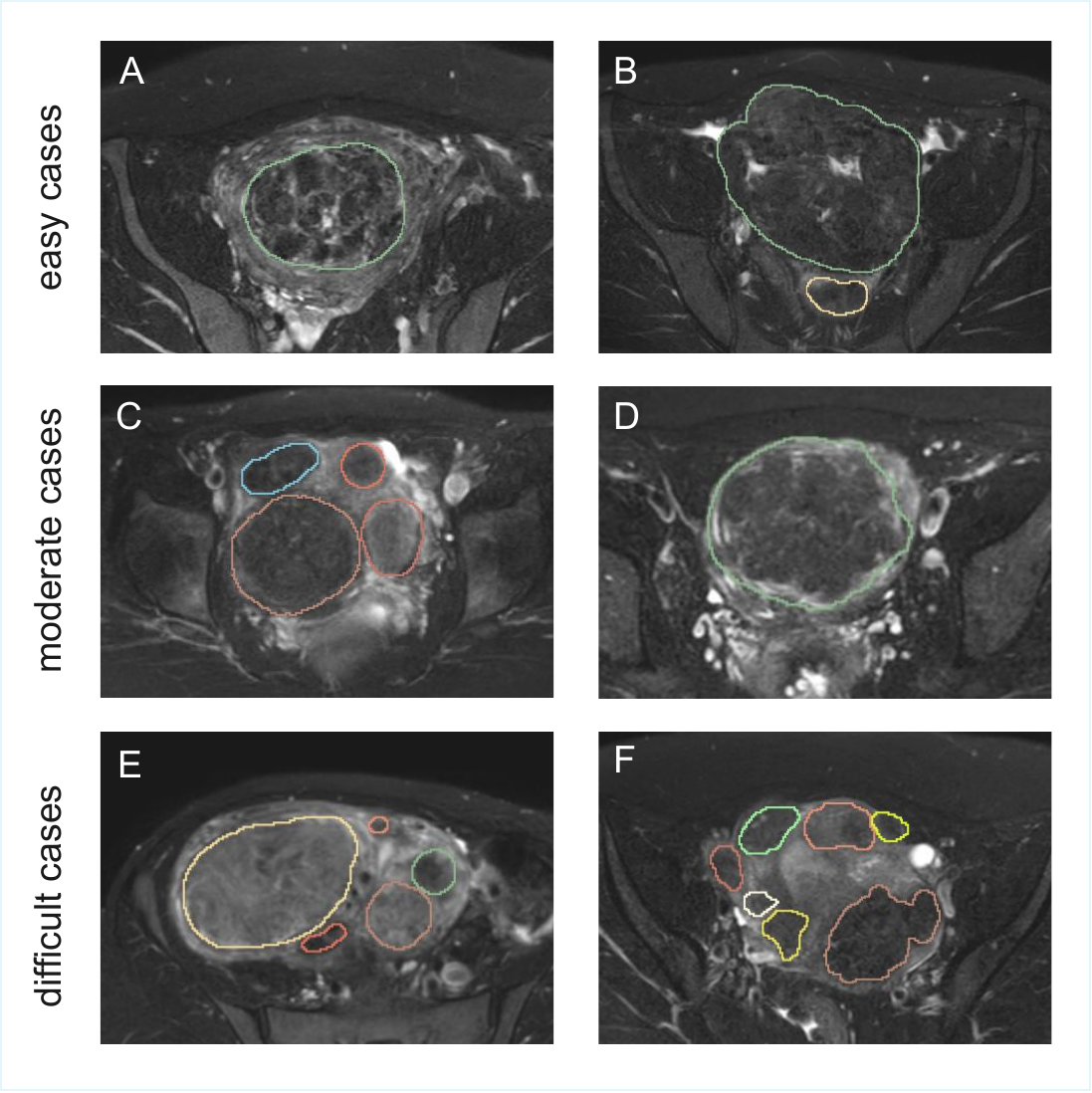}
	\caption{Test patients for fibroid segmentation neural network. For testing the fibroid segmentation neural network, six patients were selected from our dataset: two easy (Panel A and B), two moderate (Panel C and D) and two difficult cases (Panel E and F).The easy cases had a few fibroids with well defined contours, while the more difficult cases had lots of fibroids and blurred contours. The T2W SPAIR axial MRI images with the segmentation masks are shown in the images.}
	\label{fig:test_slices}
\end{figure}

\begin{figure}
	\centering
	\includegraphics[width=13 cm]
	{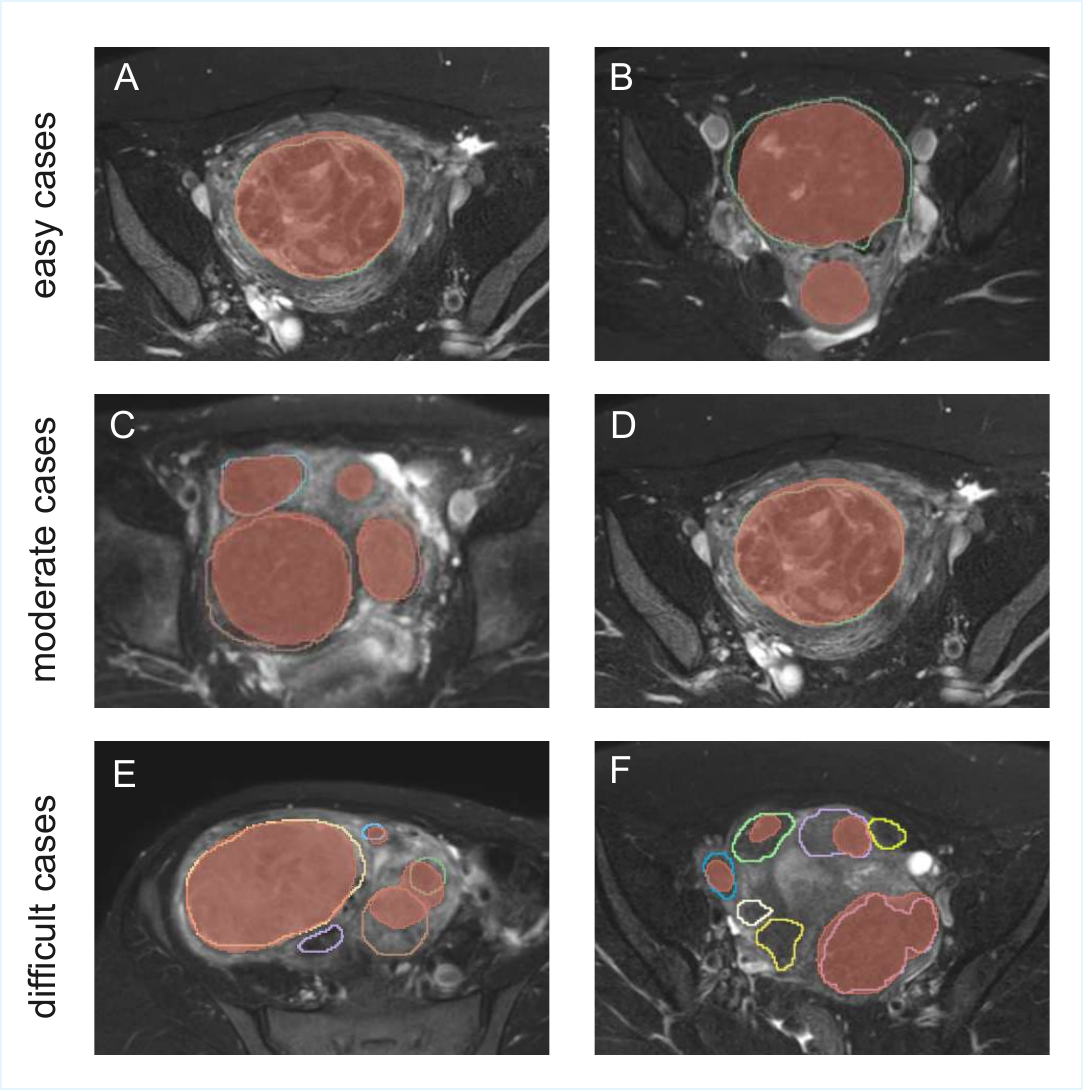}
	\caption{Result masks of the fibroid segmentation neural network. For testing the fibroid segmentation neural network, six patients were selected from our dataset: two easy (Panel A and B), two moderate (Panel C and D) and two difficult cases (Panel E and F). One slice of each patient is selected, the reference masks are show with contours, the test masks are shown in red filled areas. The T2W SPAIR axial MRI images with the segmentation masks are shown in the images.}
	\label{fig:test_masks}
\end{figure}

The fine tuning of the hyperparameters can adjust the MSI to different clinical application. We applied \texttt{pytorch} \citep{pytorch} package to train a neural network for segmentation. Using the resulting network, we segmented six images. The test images were manually selected: two easy cases, two moderate cases and two difficult cases (see Fig. \ref{fig:test_slices}). The easy cases (Fig. A. and Fig. B.) had one or two fibroids with well defined contours. The moderate cases (Fig. C. and Fig. D.) had more fibroids or the contours of the fibroids were not well defined. The difficult cases (Fig. E. and Fig. F.) had lots of fibroids or the contours of the fibroids were shaded and blurred so the delineation is harder in these two cases. One slice of all six patients is also shown in Fig. \ref{fig:test_masks}., where the reference and test masks are present.

\begin{figure}
	\centering
	\includegraphics[width=7 cm]
	{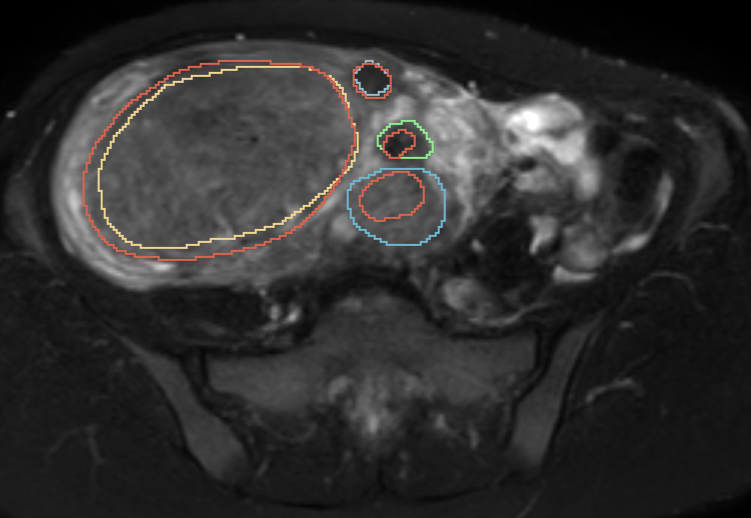}
	\caption{Representative slice with reference and test fibroid contours. The reference contours are shown with yellow, green, purple and blue colors, the test contour is shown in red. The corresponding traditional metric values (Dice and Jaccard index, average Hausdorff distance) and the MSI values with different hyperparameters are shown in Table \ref{table1}.}
	\label{fig:msi_trad_table}
\end{figure}

The MSI is able to adapt to different clinical applications by the modification of the $\texttt{il}$ and $\texttt{ol}$ hyperparameters. A representative example is shown in Fig. \ref{fig:msi_trad_table}. The MSI was calculated with the default $\texttt{il}=1$, $\texttt{ol}=1$ hyperparameters, as well as with $\texttt{il}=5, 10$ and $\texttt{ol}=5, 10$. The traditional metric values are also indicated. If there are no special needs, the value of 0.857 could be clinically relevant, which is in agreement with the traditional metric values. In contrast, if we would want to measure the volumes of the fibroids for treatment, the inner deviation had more severe consequence. That is why the segmentation of this slice should have a low score (0.512 or 0.392). If the outer deviation would have more clinical impact, MSI value of 0.733 or 0.626 could be achieved. The MSI values with $\texttt{ol}=5, 10$ are still larger than the MSI values with $\texttt{il}=5, 10$, because the segmentation has more inner alterations than outer.

\begin{table}[t]
	\centering
	\begin{tabular}{c c c c c c c}
		\hline
		MSI & contour1 & contour2 & contour3 & contour4 & \texttt{il} & \texttt{ol}\\ 
		\hline
		0.858 & 0.813 & 0.906 & 0.902 & 0.759 & 1 & 1 \\
		0.734 & 0.813 & 0.906 & 0.178 & 0.654 & 1 & 5 \\
		0.626 & 0.813 & 0.906 & 0.037 & 0.440 & 1 & 10 \\  
		0.512 & 0.117 & 0.278 & 0.898 & 0.746 & 5 & 1 \\
		0.392 & 0.012 & 0.070 & 0.897 & 0.713 & 10 & 1 \\
		\hline  
	\end{tabular}
	\caption{MSI values with different \texttt{il} and \texttt{ol} hyperparameters for the representative slice of Fig. \ref{fig:msi_trad_table}. The first column shows the MSI values for the representative slice. In the contour1, 2, 3, 4 columns, the MSI values for the individual contours are indicated. The \texttt{il}=1, 5, 10 and \texttt{ol}=1, 5, 10 was used.}\label{table1}
\end{table}

\subsection{MSI can be adapted to clinical applications: an example with prostate segmentation}

\begin{figure}
	\centering
	\includegraphics[width=13 cm]
	{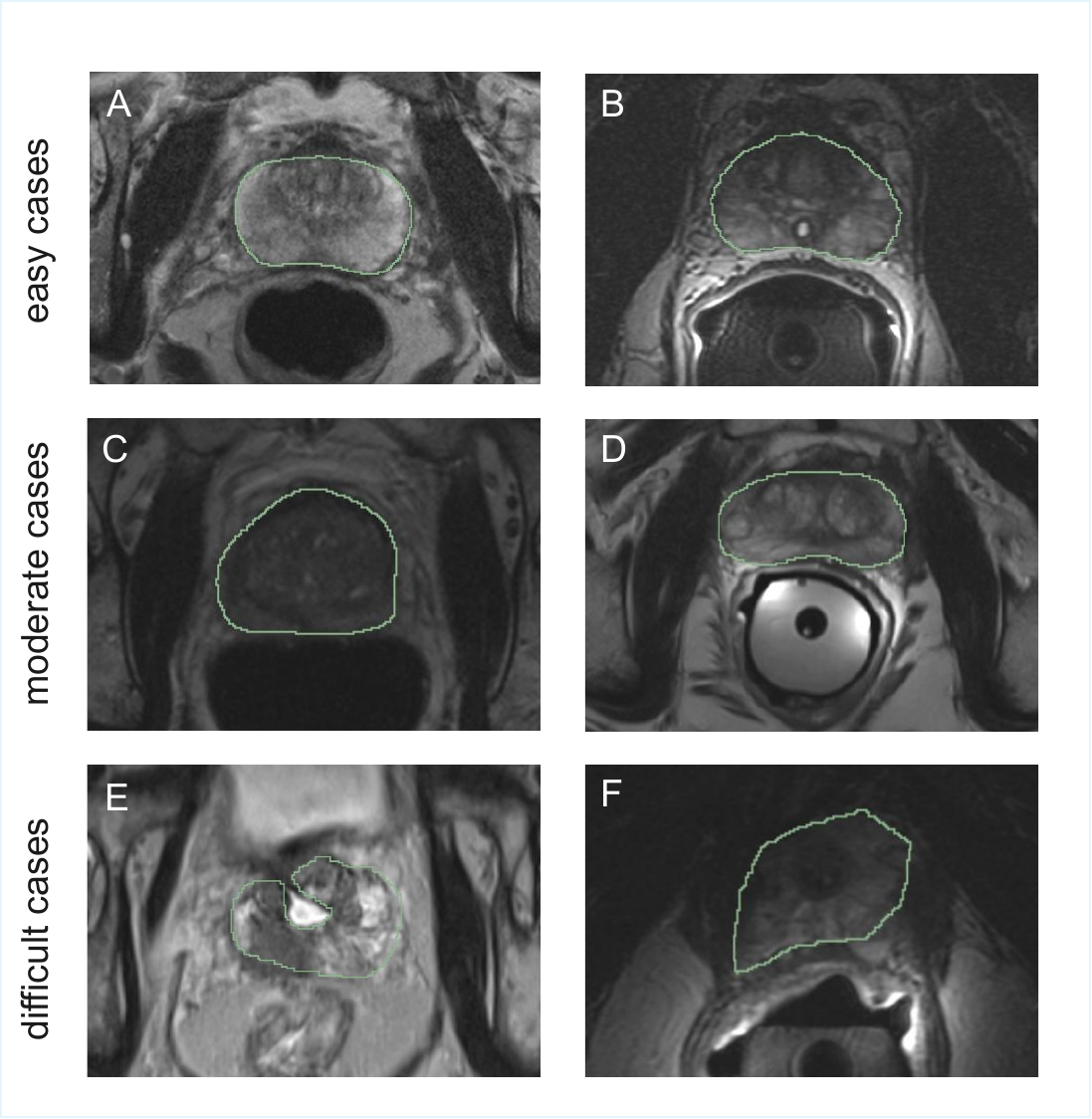}
	\caption{Test patients for prostate segmentation neural network. For testing the prostate segmentation neural network, six patients were selected from our dataset: two easy (Panel A and B), two moderate (Panel C and D) and two difficult cases (Panel E and F).The easy cases had well defined boundaries, while the more difficult cases had blurred contours, irregular shape or worse image quality. The T2W SPAIR axial MRI images with the segmentation masks are shown in the images.}
	\label{fig:test_prostate}
\end{figure}

For the demonstration of the adaptability and clinical usefulness of the pipeline, a prostate anatomic segmentation dataset was chosen. We created a segmentation neural network to generate test masks for the 6 selected test patients (Fig. \ref{fig:test_prostate}.). 

\begin{figure}
	\centering
	\includegraphics[width=13 cm]
	{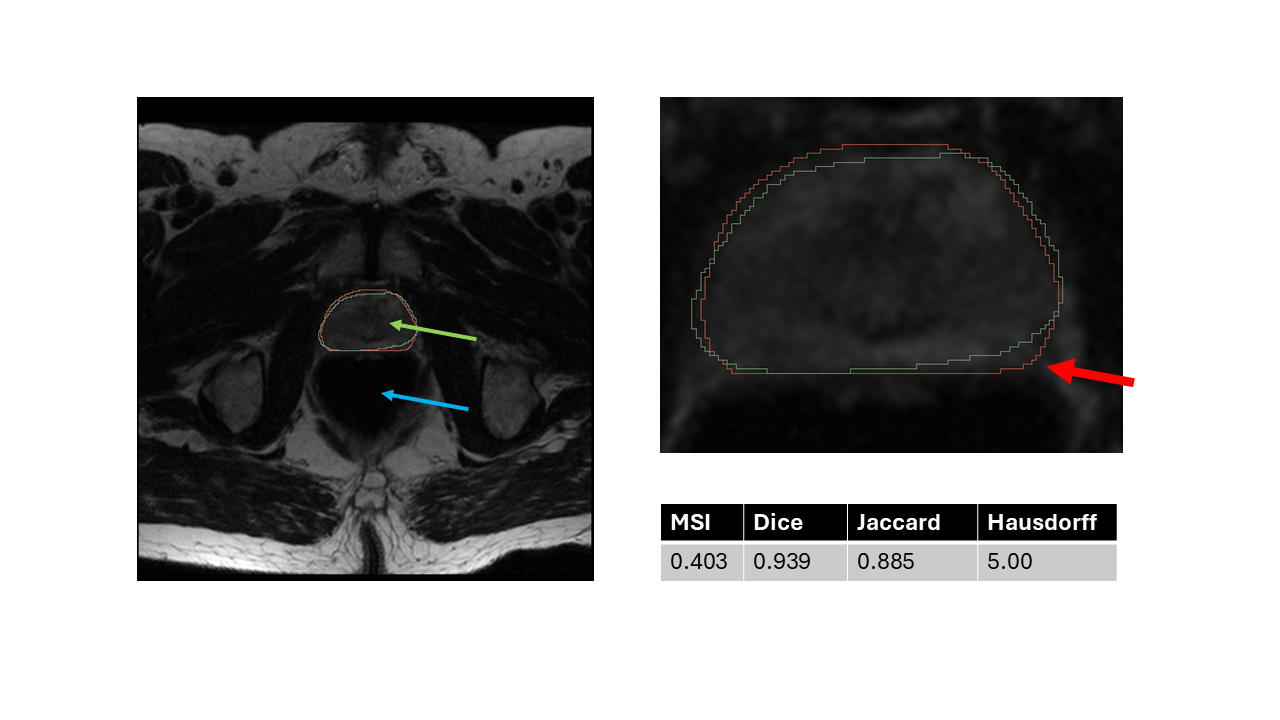}
	\caption{Representative test slice with crucial outer deviation of the test segmentation. The reference segmentation is show in green, the neural network predicted segmentation is shown in red. The predicted segmentation has an outer deviation, which reaches in the urinary bladder. In case of radiotherapy, this deviation would have serious effects, so the segmentation is unacceptable, which is only indicated by the MSI metric (all the traditional metrics show good values). The MSI value was calculated using $\texttt{il}=1$, $\texttt{ol}=10$ hyperparameters. The urinary bladder is indicated by green arrow, the prostate is indicated by blue arrow in the T2W MRI images.}
	\label{fig:outer}
\end{figure}

The impact of the outer deviation is crucial in the current clinical application, that is why we need the desired metric to represent the segmentation defects which lays out of the reference segmentation. This phenomenon can be nicely studied in the following case. In Fig. \ref{fig:outer} we represent one slice of one test segmentation, where the outer segmentation deviation reaches the urinary bladder. In this case the segmentation is unacceptable. The traditional metrics and the value of MSI with $\texttt{il}=1$, $\texttt{ol}=10$. The Dice (0.939) and Jaccard score (0.886) show high values, the Hausdorff distance show a medium value (5.0), while the MSI (0.403) has a very low value. Only the MSI characterizes the segmentation correctly. The metric values considering all slices and the masks are presented in Appendix (see Fig. \ref{fig:ax_metrics}).

\section{Discussion}

Prostate cancer is the second most common cancer worldwide. During their lifetime, approximately 1 out of 8 men will be diagnosed with prostate cancer. The estimation for prostate cancer for 2025 by the American Cancer Society predicts about 313,780 new cases and about 35,770 deaths \citep{ACS2025}.
The segmentation of the prostate in MRI images has an important role not only in radiotherapy treatment planning and follow up, but in other diagnostic procedures, such as PSA (prostate-specific antigen) density or tumor/prostate ratio calculation \citep{Jimenez2023}. What is more, for multi-model imaging, the registration algorithm needs the prostate segmentation for initialization \citep{litjens2014}.

Uterine myomas - also known as leiomyomas or fibroids - are caused by the abnormal growth of the uterus' smooth muscle cells. They are the most common benign tumor of the female reproductive system, affecting 20-30\% of the female population \citep{li2023}. Although at least 50\% of the fibroids are asymptomatic \citep{divakar2008}, 5-10\% of the cases are associated with infertility \citep{ahmad2023} and it is the most common indication for hysterectomy worldwide \citep{pan2023}.
The clinical treatment of the fibroids reqiure the consideration of the location of the fibroid(s) inside the uterus, the relationship with the uterine wall and cavity. For this reason, the automatic diagnosis and preoperative evaluation requires the segmentation of fibroids, uterine wall and cavity in MR images \citep{pan2023}.

Although there are some reviews considering the evaluation metrics for general medical image segmentation \citep{park2024} \citep{muller2022}, as well as for special application areas such as blood vessel segmentation \citep{Moccia2018}, retinal optical coherence tomography \citep{Zhang2025} and radiotherapy \citep{Mackay2023}. The evaluation of the segmentation algorithms impacts the optimization of the segmentation algorithms as it directly influence how the performance is measured and compared \citep{Wei2024}. However, currently standardized and clinically relevant evaluation protocol is not available. There are an emerging number of automatic methods, as they may speed up the segmentation process. However, there is no reliable model performance assessment method, and statistical bias is also reported caused by the incorrect metric implementation or usage \citep{muller2022}. It is shown that common quantification metrics do not reflect clinical accceptance in case of heart contouring from CT images \citep{vandenOever2022}. Determining the most appropriate evaluation measures is a very challenging task, this is why it it is essential to match the possible metrics to the segmentation objectives \citep{fenster2005}.

We provide an adaptable pipeline for medical image segmentation evaluation tasks, our code provides a skeleton for further refinement for other specific applications.

There is option for hyperparameter optimalization, i.e. the best choice for \texttt{il} and \texttt{ol} levels could be selected. Of course the best choice depends on the clinical use - if the inside or outside deviation have more important consequence.

In some cases - such as the prostate - there is only one mask in one image slice. If there are more than one contours in an image slice - i.e. in the case of fibroids --, pairing the reference and test contours is needed for evaluation. We used a simple method, the algorithm finds the closest center of mass for each test contours. However, this process can be improved and modified according the current clinical use. There are many more sophisticated algorithms for such tasks, but our aim was to provide one simple solution and let the modifications designed for the different tasks.
This pipeline requires manual intervention if the number of reference and test masks not equals. By default, these kind of slices receive MSI score of zero, as these segmentations are unacceptable. However, more sophisticated separation of these slices could be useful, as there may be better and worse segmentations as well in this group.

What is more, we get one MSI value for each contour and these scores are aggregated to one final MSI value for each slice. We used the median for this aggregation, but it can be further discussed if other methods may improve the results. Furthermore, giving one MSI score for one patient may also be useful for the isolation of very good or unacceptable segmentations. These changes can be easily made in our pipeline.

\section{Conclusion}

It can be concluded that we have developed an easy to use, adaptable pipeline for medical segmentation tasks. The pipeline facilitates comprehensive performance analysis by computing not only traditional segmentation metrics — such as Dice and Jaccard scores and the average Hausdorff distance — but also introduces the Medical Similarity Index to assess segmentation agreement with enhanced clinical relevance. We provide an outline for each image processing step. The code is available and follows sustainable, object-oriented design principles. This allows for straightforward customization and extension to suit specific research or clinical needs.

However, the current implementation has several limitations. The pipeline is designed to operate with NIfTI-formatted input images and segmentation masks, which may limit its applicability to datasets in other formats that require prior conversion. Additionally, images containing multiple segmentation masks present challenges for contour pairing. The pipeline identifies the problematic slices where the number of reference and test masks is not equal, or the contour pairing cannot be performed based on the closest center of mass method (as shown in \ref{fig:pairing_exc}). While this method provides a good starting point and enables the pipeline to function, it remains relatively simple. More advanced contour pairing techniques could improve robustness, especially in complex segmentation configurations. Such enhancement must be developed with consideration for the specific requirements of the desired clinical application.

\backmatter


\section*{Declarations}

\begin{itemize}

\item Ethics approval and consent to participate

This study was approved by the Institutional Review Board (Semmelweis University Regional and Institutional Committee of Science and Research Ethics, SE-RKEB: 172/2022). As this was a retrospective study, the need for written informed patient consent was waived by the ethics committee. All procedures performed in this study involving human participants were in accordance with the ethical standards of the Declaration of Helsinki. All patient data were analyzed anonymously.

\item Consent for publication

Not applicable.

\item Availability of data and materials

The codes and notebooks used in the current study are available in the cited GitHub repository \citep{szuzina_bld}. The prostate dataset is available from the cited publication, the images of fibroid dataset are not publicly available due to personal rights, however, the generated segmentation masks are available from the cited repository.

\item Competing interests

The authors declare that they have no competing interests.

\item Funding

The research reported here was supported by the National Research, Development and Innovation Office (NKFIH) in Hungary [grant number RRF-2.3.1-21-2022-00006]. Szuzina Fazekas receives a grant from the Gedeon Richter Talentum Foundation within the framework of the Gedeon Richter Excellence PhD Scholarship.

\item Author contribution

SzF: Conceptualization, Methodology, Software, Investigation, Data Curation, Visualization, Writing - Original Draft; BBK: Conceptualization, Data Curation, Writing - Review \& Editing; BV: Resources, Writing - Review \& Editing; PMH: Resources, Writing - Review \& Editing; ZsV: Conceptualization, Methodology, Software, Data Curation, Writing - Review \& Editing, Supervision
All authors read and approved the final manuscript.

\end{itemize}

\begin{appendices}

\label{appendix}

\section*{Google Colaboratory notebook}

The notebook's preparation part consists of the necessary imports and cloning of the GitHub repository. The URL of the reference and test files must be provided for the downloading process.

In the MSI calculation part, the testing calculations paragraph inputs the number of the current patient and current slice, and the inside and outside penalty levels can be declared; the algorithm will give one MSI value as the final result. The Visualization paragraph features various images and graphs to facilitate understanding and experimentation with the dataset and MSI values. The Split masks paragraph provides an opportunity to handle the concave mask case, where two touching masks are drawn together (see \ref{split}).

\section*{Figures}

\appendix
\renewcommand{\thefigure}{A\arabic{figure}}
\setcounter{figure}{0}

\begin{figure}[htp]
	\centering
	\includegraphics[width=13 cm]
	{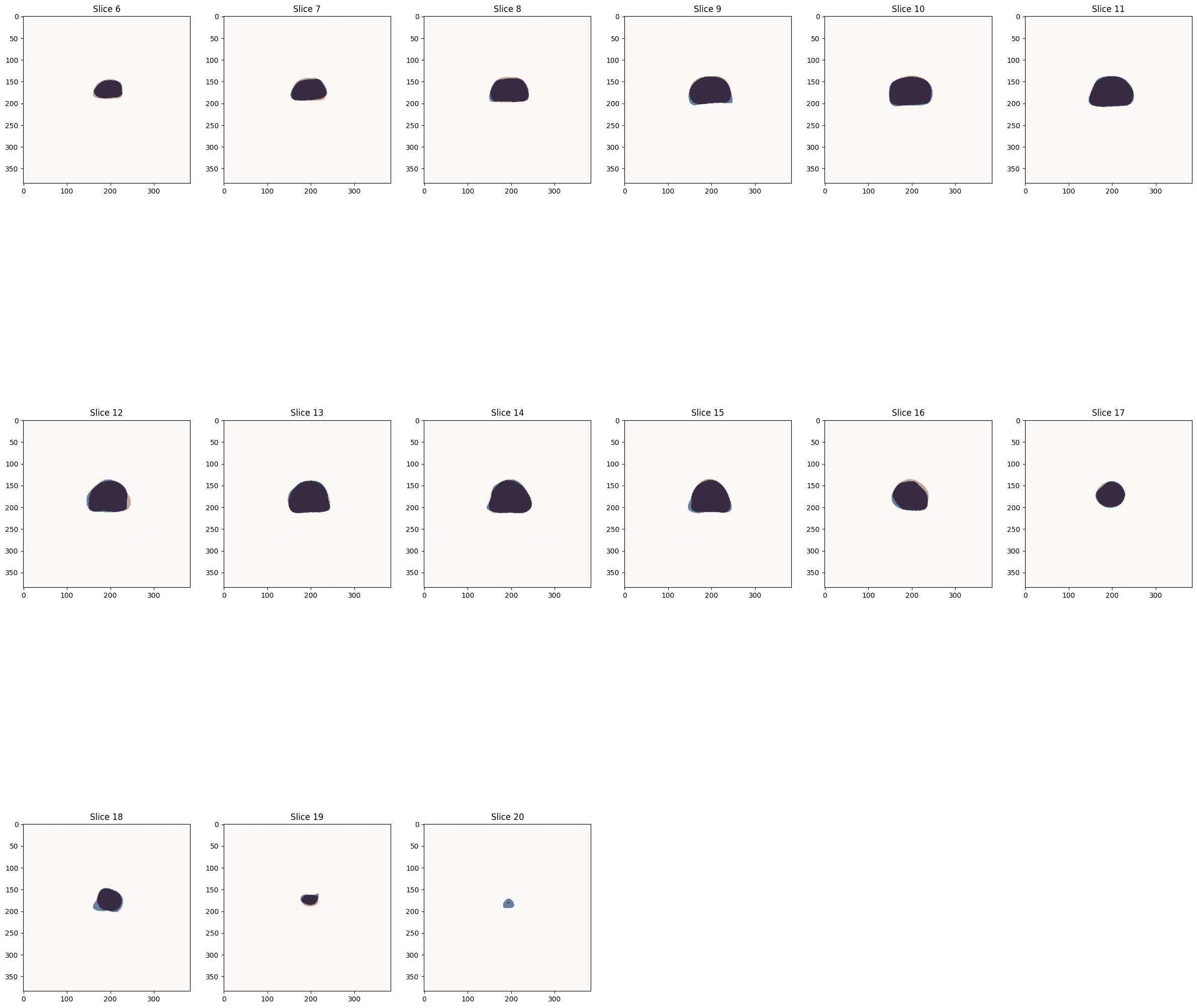}
	\caption{The metric values and the segmentation masks of the representative patient. The reference mask is shown in blue, the test masks are shown in orange. The MSI values were calculated with $\texttt{il}=1$, $\texttt{ol}=10$ hyperparameters, the values are shown in Table \ref{table2}.}
	\label{fig:ax_metrics}
\end{figure}

\begin{table}[htp]
	\centering
	\begin{tabular}{c c c c c}
		\hline
		slice index & MSI & Dice & Jaccard & Hausdorff\\
		\hline
		6 & 0.205179 & 0.922558 & 0.856248 & 5.385165\\
		7 & 0.403701 & 0.939892 & 0.886601 & 5.000000\\
		8 & 0.416621 & 0.951325 & 0.907168 & 5.656854\\  
		9 & 0.686597 & 0.949580 & 0.904000 & 8.485281\\
		10 & 0.727526 & 0.961066 & 0.925051 & 4.000000\\
		11 & 0.619143 & 0.977774 & 0.956514 & 3.000000\\
		12 & 0.730977 & 0.937705 & 0.882716 & 9.000000\\
		13 & 0.596483 & 0.972026 & 0.945574 & 3.605551\\
		14 & 0.767887 & 0.970246 & 0.942211 & 5.385165\\
		15 & 0.671288 & 0.953519 & 0.911167 & 7.280110\\
		16 & 0.254589 & 0.922909 & 0.856854 & 7.071068\\
		17 & 0.732791 & 0.957405 & 0.918290 & 4.242641\\
		18 & 0.634041 & 0.895528 & 0.810820 & 13.892444\\
		19 & 0.551165 & 0.835125 & 0.716923 & 5.830952\\
		20 & 0.567933 & 0.048998 & 0.025114 & 14.212670\\
		\hline  
	\end{tabular}
	\caption{The MSI values for the slices shown in Fig. \ref{fig:ax_metrics}. The MSI values were calculated with $\texttt{il}=1$, $\texttt{ol}=10$ hyperparameters.}\label{table2}
\end{table}

\end{appendices}

\end{document}